\documentclass[prl,superscriptaddress,showpacs,twocolumn]{revtex4-1}

\usepackage{graphicx}
\usepackage{color}
\usepackage{verbatim}
\usepackage{subfigure}
\usepackage{amssymb}
\usepackage{amsmath}

\begin{document}
	
\title{Theory for Glassy Behavior of Supercooled Liquid Mixtures}
\author{Shachi Katira}
\affiliation{
Department of Chemistry, University of California, Berkeley, CA, USA}
\affiliation{
Department of Chemical and Biomolecular Engineering, University of California, Berkeley, CA, USA}

\author{Juan P. Garrahan}
\affiliation{School of Physics and Astronomy, University of Nottingham, Nottingham, NG7 2RD, UK}
\affiliation{Centre for the Mathematics and Theoretical Physics of Quantum Non-Equilibrium Systems, University of Nottingham, Nottingham, NG7 2RD, UK}

\author{Kranthi K. Mandadapu}
\affiliation{
Department of Chemical and Biomolecular Engineering, University of California, Berkeley, CA, USA}
\affiliation{Chemical Sciences Division, Lawrence Berkeley National Laboratory, Berkeley, CA, USA}
\begin{abstract}
We present a model for glassy dynamics in supercooled liquid mixtures. Given the relaxation behavior of individual supercooled liquids, the model predicts the relaxation times of their mixtures as temperature is decreased. The model is based on dynamical facilitation theory for glassy dynamics, which provides a physical basis for relaxation and vitrification of a supercooled liquid. This is in contrast to empirical linear interpolations such as the Gordon--Taylor equation typically used to predict glass transition temperatures of liquid mixtures.  To understand the behavior of supercooled liquid mixtures we consider a multi-component variant of the kinetically constrained East model in which components have a different energy scale and can also diffuse when locally mobile regions, i.e., excitations, are present. Using a variational approach we determine an effective single component model with a single effective energy scale that best approximates a mixture. When scaled by this single effective energy, we show that experimental relaxation times of many liquid mixtures all collapse onto the `parabolic law' predicted by dynamical facilitation theory. The model can be used to predict transport properties and glass transition temperatures of mixtures of glassy materials, with implications in atmospheric chemistry, biology, and pharmaceuticals.

\end{abstract}
\maketitle
\noindent

This paper presents a model for glassy dynamics in mixtures of liquids. Understanding and predicting glassy behavior of liquid mixtures and solutions has been important in investigating the fundamental properties of glassy dynamics and the glass transition~\cite{blochowicz2004beta,wang2010structural,foffi2003mixing}, as well as in areas of research where solutions of liquids undergo vitrification into amorphous solids, and where crystallization needs to be avoided. These areas include chemistry of solutions~\cite{angell1978glass}, and pharmaceutical and food industries~\cite{hancock1997characteristics,yu2001amorphous,bhandari1999implication}. Insight into properties of glassy mixtures is also important in more complex systems such as atmospheric organic aerosols that show glassy behaviour~\cite{virtanen2010amorphous,vaden2010morphology,cappa2011evolution,koop2011glass}, and in the preservation of biological cells where vitrification may assist the recovery of cells after freezing and/or desiccation~\cite{crowe1998role,green1989phase}. The typical approach to predict the glass transition temperature of liquid mixtures is an empirical linear interpolation between the glass transition temperatures of individual components, such as the Gordon--Taylor equation~\cite{gordon1952ideal}, without a physical basis in a microscopic theory for glassy dynamics. In this work we use the perspective of dynamical facilitation theory to develop a model for glassy dynamics of binary mixtures, which can be used to predict relaxation behavior and glass transition temperatures of supercooled liquid mixtures.

When a glass-former is cooled below its onset temperature ($T^\mathrm{o}$) its relaxation time exhibits a super-Arrhenius increase with decreasing temperature~\cite{ediger1996supercooled,angell2000relaxation}, and its microstructure exhibits dynamical heterogeneity, i.e., transient but distinct mesoscopic regions of particle mobility and immobility. The increase in relaxation time is explained by dynamical facilitation theory, the idea that regions of particle mobility (`excitations') \emph{facilitate} the motion of neighboring regions in a \emph{hierarchical} manner, with small motions leading to larger motions~\cite{palmer1984models,sollich1999glassy,sollich2003glassy,garrahan2002geometrical,chandler2010dynamics,keys2011,speck2019dynamic}. The theory predicts a super-Arrhenius form, quadratic in inverse temperature, for the increase in relaxation time $\tau$ with decreasing temperature. This quadratic form collapses experimental data for relaxation times of a wide array of glass-forming liquids onto one universal curve, the parabolic law, given by~\cite{speck2019dynamic,elmatad2009corresponding}
\begin{equation}\label{parabolic}
\ln \tau/\tau^\mathrm{o} \sim J^2 (\beta - \beta^\mathrm{o})^2.
\end{equation}
Here $\beta=1/T$ is inverse temperature, $\beta^\mathrm{o} = 1/T^\mathrm{o}$ is the inverse onset temperature, $J$ is a property of the liquid and is related to the free energy of creating an excitation, and $\tau^\mathrm{o}$ is the relaxation time at $T^\mathrm{o}$. With knowledge of the onset temperature $T^\mathrm{o}$, the energy scale $J$, and the reference time scale $\tau^\mathrm{o}$, one can predict the relaxation time at a given temperature, and the glass transition temperature for a given cooling rate~\cite{keys2013calorimetric,hudson2018nature}. The parabolic law (Eq.~\ref{parabolic}) differs from the commonly used empirical Vogel--Fulcher--Tammann relation~\cite{ediger1996supercooled}, $\ln \tau \sim \mathrm{const}/(T-T_\mathrm{K})$, and other proposed theories~\cite{adam1965temperature,gotze1992relaxation,mezard1999thermodynamics,xia2001microscopic}, in that it does not show a singularity at non-zero temperatures. We note that there is no microscopic physical theory to understand the onset temperature, which is the temperature at which the relaxation behavior crosses over from an Arrhenius to a super-Arrhenius form.
The idea of dynamical facilitation is based on prototypical models with kinetic constraints~\cite{Ritort2003,Garrahan2011,Garrahan2018}, such as the East model~\cite{jackle1991hierarchically}, that exhibit hierarchical glassy relaxation. The East model is a one-dimensional lattice of $N$ spins with variables $n_i=0,1$, where $i=1,2\ldots N$. Each lattice site represents a region of particles and the spin variable, $n_i$, indicates the mobility of the region.  $n_i=1$ represents an excitation, i.e., a region of mobile particles, and $n_i=0$ represents a region of immobile particles. The model has a noninteracting energy function $H_\mathrm{East}=J \sum_i n_i$, where $J$ represents the energy required to generate particle mobility, i.e., an excitation, in a particular region. 
The dynamics proceeds via single spin flips with the constraint that a site can flip only if its left neighbor is \emph{excited}, i.e., $n_{i-1}=1$, based on the idea that particle mobility is \emph{facilitated} in the vicinity of other mobile particles~\cite{garrahan2002geometrical,chandler2010dynamics,keys2011}. The typical concentration of excitations is $c=\langle n \rangle = \exp(-\beta J)/(1+\exp(-\beta J)) \sim \exp(-\beta J)$ at low temperatures. 

Relaxation of the system is defined as all of the spins changing their state at least once, and progresses in the direction of facilitation, i.e., from left to right. Sollich and Evans~\cite{sollich1999glassy,sollich2003glassy} rationalized the timescale of relaxation by considering domains between excitations, e.g.\,1000001. 
The \emph{height} of the energy barrier to relaxation is defined as the maximum number of excitations required in a single configuration during the process of relaxing, or flipping, the rightmost excitation. The minimum height of the barrier for a domain of length $ l = 2^k$ is reasoned to be $k$ excitations by iteratively bisecting the domain. For example, a domain of length $4L$ requires twice as many excitations as a domain of length $2L$, which in turn requires twice as many excitations as a domain of length $L$. The minimum energy barrier is therefore $kJ = (\ln l/\ln 2) J.$ For the average equilibrium domain length $l^\mathrm{eq} = 1/c \sim \exp(\beta J)$, the energy barrier is $\beta J^2/\ln 2$. The relaxation time $\tau$ is inversely related to the Boltzmann probability of this energy barrier, giving $\tau \sim \exp(\gamma \beta^2 J^2)$, which is the parabolic law. The factor $\gamma$ represents the number of paths available to relax a domain, which are not accounted for in the Sollich--Evans argument, and more rigorous analysis shows that it is bounded as $1/2\ln2 < \gamma < 1/\ln2$~\cite{chleboun2013time}.
\newline

{\bf \em A Multi-Component East Model.} In the presence of more than one component, the relaxation dynamics includes not only dynamical heterogeneity but also component diffusion and mixing. The interplay between heterogeneous dynamics and mixing in supercooled liquid mixtures is not understood. With the perspective of dynamical facilitation theory and the East model we construct a multicomponent lattice model for mixtures of glass-formers that are well-mixed and do not phase separate. We begin by considering a mixture of two components with different energy scales, $J_0$ and $J_1$. Each lattice site represents a region of molecules of one component or the other, represented by the lattice variables $p_i=0,1$.  The energy cost for an excitation depends on the type of component, i.e., the value of $p_i$ for that site. The Hamiltonian for this two component system is 
\begin{equation}\label{h}
H(\{n_i,p_i\},x) = \sum_i^N n_i (p_i J_1 + (1-p_i)J_0) + C(\{p_i\}),
\end{equation}
where $J_0$ and $J_1$ are the energies required to excite a spin with component $p_i=0$ or 1 respectively. Here $C(\{p_i\})$ is a constraint function that ensures that the total number of each component remains fixed:
\[C(\{p_i\})
    = 
\begin{cases}
    0,& \text{if } \frac{1}{N}\sum_i^N p_i = x\\
    \infty,              & \text{otherwise.}
\end{cases}
\]
where $x$ is the fraction of sites with $p_i=1$.
\newline
\begin{figure}[t] 
\centering
\vspace{0.2in}
\textcolor{red}{1}\textcolor{blue}{0}\textcolor{red}{0}\textcolor{red}{0}\textcolor{blue}{0}\textcolor{blue}{0}\textcolor{blue}{0}\textcolor{blue}{0}\textcolor{red}{0}\textcolor{blue}{0}\textcolor{red}{1} $\rightarrow$
\underline{\textcolor{red}{1}\textcolor{blue}{1}}\textcolor{red}{0}\textcolor{red}{0}\textcolor{blue}{0}\textcolor{blue}{0}\textcolor{blue}{0}\textcolor{blue}{0}\textcolor{red}{0}\textcolor{blue}{0}\textcolor{red}{1} $\rightarrow$ 
\underline{\textcolor{blue}{1}\textcolor{red}{1}}\textcolor{red}{0}\textcolor{red}{0}\textcolor{blue}{0}\textcolor{blue}{0}\textcolor{blue}{0}\textcolor{blue}{0}\textcolor{red}{0}\textcolor{blue}{0}\textcolor{red}{1}
\vspace{0.1in}
\caption{Schematic of a two component East lattice with colors representing the component variables $p_i$. A spin can only change its value if its left neighbor is excited. Two adjacent excited spins can exchange colors, i.e, their component variables, resulting in diffusion (underlined).}
\label{fig:colorspins}
\end{figure}

Extending the idea of dynamical facilitation theory that regions of mobile particles facilitate the movement of neighboring immobile regions, irrespective of the type of particles in each region, we define the dynamics of the mixture model as similar to the single component East model, i.e., a spin can flip only if its left neighbor is excited. The rates of creating and destroying an excitation at site $i$ are
\begin{equation} 
\begin{split}
k_{0\rightarrow1,i} &\propto (p_i\, e^{-\beta J_1} + (1-p_i)\, e^{-\beta J_0}) n_{i-1}\\
k_{1\rightarrow0,i} &\propto n_{i-1},
\end{split}
\end{equation}
where $n_i=0,1$ again represents regions of immobile and mobile particles respectively. With the reasoning that only regions of mobile particles are able to diffuse, two adjacent regions will be able to exchange components only if they are mobile. We model this as two adjacent spins exchanging $p_i$ variables only if they are both excited, as illustrated in Fig.~\ref{fig:colorspins}. This occurs with rate
\begin{equation}
r_{i,i\pm1} \sim n_{i}\, n_{i \pm 1} ( (1-p_i) p_{i\pm1} + (1-p_{i\pm1}) p_i ),
\end{equation}
and is based on the assumption that mixing of particles in two adjacent regions occurs over the same timescale as motion of particles.

\begin{figure*}[ht] 
\includegraphics[scale=0.32]{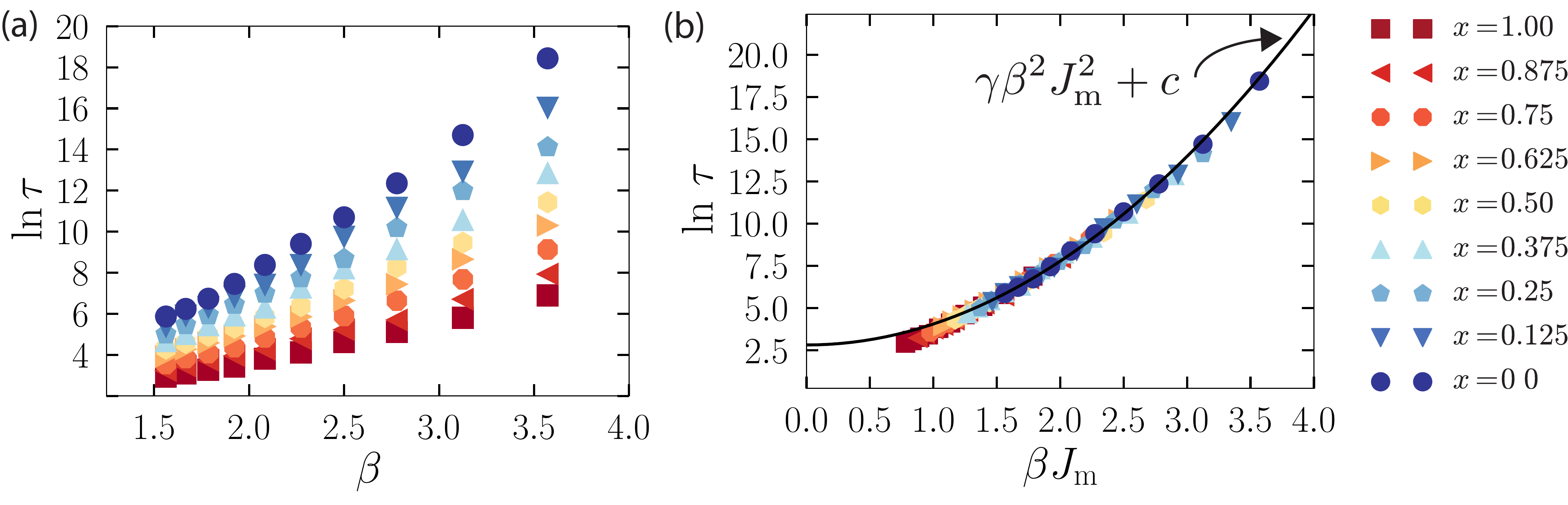}
\caption{(a) Relaxation times $\tau$ of East model mixtures of two components $J_1=0.5$ and $J_0$=1.0 as a function of inverse temperature $\beta$. $x$ is the fraction of spins of type 1. (b) All the curves collapse onto the parabolic law when the inverse temperature axis is scaled by the effective energy $J_\mathrm{m}=xJ_1+(1-x)J_0$. The factor $\gamma=1.18$ within the predicted bounds $1/2\ln2 < \gamma < 1/\ln2$~\cite{chleboun2013time}, and $c$ is an additive constant. Error is smaller than the symbols used.}
\label{fig:all_east}
\end{figure*}

To illustrate the relaxation behavior of the mixture model we choose a system with two energy scales, $J_0=1$ and $J_1=0.5$. We define the relaxation time, $\tau$, as the time taken for ninety percent of the system to change its spin variable, $n_i$, at least once, which represents ninety percent of particles making non-trivial displacements at least once~\cite{keys2011}. In Fig.~\ref{fig:all_east}(a), we show relaxation times for mixtures with different fractions of the two components. Similar to a single component system, the curves all appear super-Arrhenius, with the mixture relaxation curves lying in between the relaxation curves for the individual components.
\newline

{\bf \em An Effective Single Component Model.} Based on the super-Arrhenius relaxation behavior of the mixtures in Fig.~\ref{fig:all_east}(a)
we hypothesize that a mixture effectively behaves as a single, homogeneous material with an effective, predictable energy scale for excitations. This means that mixing of components gives rise to a new, effective dynamical heterogeneity at different length and time scales compared to the original components. To capture the effective behavior of the two component mixture we postulate a single component East Hamiltonian, $H_\mathrm{m}$, with an effective energy scale $J_\mathrm{m}$. We attempt to predict $J_\mathrm{m}$ as a function of the energy scales of the individual component materials. The postulated single component Hamiltonian $H_\mathrm{m}$ is given by 
\begin{equation}
H_\mathrm{m}(\{n_i,p_i\},x) = J_\mathrm{m}\sum_i^N n_i + C(\{p_i\}).
\end{equation}
We use a variational method to calculate $J_\mathrm{m}$ that best approximates the partition function for the multi-component Hamiltonian $H$ (Eq.~\ref{h}). Using Jensen's inequality (sometimes known as the Gibbs--Bogoliubov--Feynman approach~\cite{feynman2018statistical,chandler1987introduction}) we have
\begin{equation}
\begin{split}
Z &= \frac{Z_\mathrm{m}} {Z_\mathrm{m}} \sum_{n_1} \ldots \sum_{n_N} \sum_{p_1} \ldots \sum_{p_N} e^{-\beta H} e^{-\beta H_\mathrm{m}} e^{+\beta H_\mathrm{m}} \\
& = Z_\mathrm{m} \langle\exp(-\beta\Delta H) \rangle_{H_\mathrm{m}} \geq Z_\mathrm{m} \exp(-\beta\langle \Delta H \rangle_{H_\mathrm{m}}),
\end{split}
\end{equation}
where $Z$ and $Z_\mathrm{m}$ are the canonical partition functions for Hamiltonians $H$ and $H_\mathrm{m}$ respectively, $\langle\ldots\rangle_{H_\mathrm{m}}$ indicates the canonical ensemble average with energy $H_\mathrm{m}$, and $\Delta H = H - H_\mathrm{m}$.

\begin{figure*}[t] \includegraphics[scale=0.365]{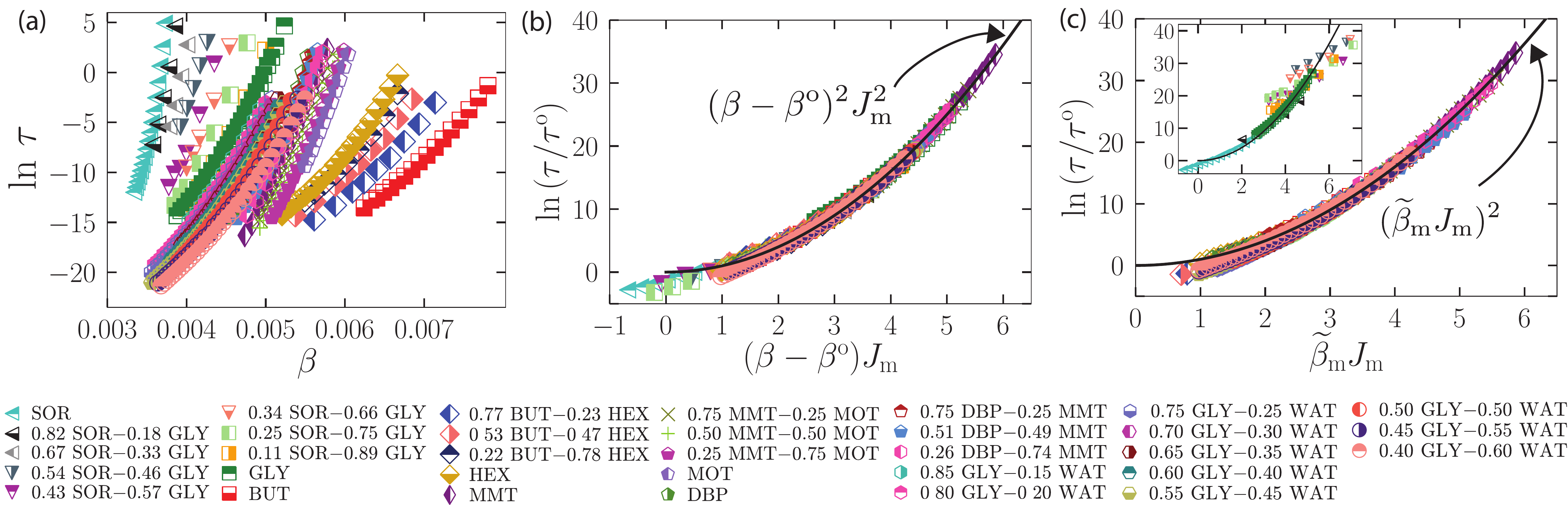}
\caption{(a) Relaxation times as a function of inverse temperature of mixtures of five pairs of liquids: sorbitol (SOR)--glycerol (GLY)~\cite{duvvuri2004binary}, 2-methyl-1-butanol (BUT)--2-ethyl-1-hexanol (HEX)~\cite{wang2005ideal}, methyl-\emph{m}-toluate (MMT)--methyl-\emph{o}-toluate (MOT)~\cite{wang2010structural},  di-\emph{n}-butyl phthalate (DBP)--methyl-\emph{m}-toluate (MMT)~\cite{wang2010structural}, and glycerol (GLY)--water (WAT)~\cite{puzenko2005relaxation}. (b) All single component and mixture curves collapse onto the parabolic law when the temperature axis is scaled by the effective mixture energy scale $J_\mathrm{m}$ (Eq.~\ref{linear}). For glycerol--water mixtures we use $40\%$ glycerol as one reference curve and pure glycerol as the second because the values of $\beta^\mathrm{o}$ and $J$ for pure water are difficult to obtain with accuracy~\cite{schuster2016glassy}. With this reference curve, all intermediate glycerol--water mixtures show excellent agreement with the parabolic law for mixtures. (c) All single component and mixture curves collapse onto the parabolic law when the temperature axis is scaled by $\widetilde{\beta}_\mathrm{m}J_\mathrm{m}$ (Eq.~\ref{betalinear}), except sorbitol--glycerol mixtures (inset).}
\label{fig:allcollapsed}
\end{figure*}
Calculating the value of $J_\mathrm{m}$ that maximizes the right hand side of the inequality we obtain
\begin{equation}\label{linear}
J_\mathrm{m} = x J_1 + (1-x) J_0.
\end{equation}
The effective energy scale $J_\mathrm{m}$ can be interpreted as giving rise to \emph{effective excitations} that occur with probability $\exp(-{\beta} J_\mathrm{m})$. Revisiting the Sollich and Evans argument~\cite{sollich1999glassy,sollich2003glassy}, the average equilibrium domain length between effective excitations is $l_\mathrm{m}^{\mathrm{eq}} =\exp({\beta} J_\mathrm{m})$, creating a minimum energy barrier of $\gamma J_\mathrm{m} \ln l_\mathrm{m}^{\mathrm{eq}} = \gamma{\beta}J_\mathrm{m}^2 $. This gives a relaxation time $\tau \sim \exp (\gamma{\beta}^2 J_\mathrm{m}^2)$, which we refer to as the parabolic law for mixtures. Using the effective energy $J_\mathrm{m}$ from Eq.~\ref{linear} to rescale the data in Fig.~\ref{fig:all_east}(a) we find that all the relaxation curves collapse onto a universal parabolic form (Fig.~\ref{fig:all_east}(b)).
This indicates that the two component system behaves as a single component system with an effective energy scale of $J_\mathrm{m}$, an average of the energy scales of the individual components. 
\newline

{\bf \em Comparison with Experimental Data.} 
With this understanding from the East model we attempt to validate the idea of an effective mixture energy scale with available experimental data for supercooled mixtures. We analyze dielectric relaxation measurements for binary mixtures of various glass-forming liquids~\cite{duvvuri2004binary,wang2005ideal,wang2010structural,puzenko2005relaxation}. Relaxation times for thirty three mixtures of five pairs of liquids are shown in Fig.~\ref{fig:allcollapsed}(a). We first fit parabolas to relaxation time data for single liquid components to determine $J$ values for the individual components. (We use the low temperature data for these fits because the higher temperature data may contain significant contributions from the non-supercooled liquid.) We then calculate $ J_\mathrm{m}$ for each mixture using Eq.~\ref{linear}. $\beta^\mathrm{o}$ and $\tau^\mathrm{o}$ for each mixture are treated as fitting parameters, and $\gamma$ is treated as absorbed into the $J$ value for all experimental data. On scaling the inverse temperature axis by $J_\mathrm{m}$, we find that all the single component as well as mixture data collapse onto the parabolic law (Fig.~\ref{fig:allcollapsed}(b)). Consistent with the model, it appears that mixtures of liquids effectively behave as a single liquid with an energy scale that is an average of the energy scales of the individual components as per Eq.~\ref{linear}. 

Recalling that the onset temperature is the temperature below which spatially and temporally heterogeneous dynamics and super-Arrhenius relaxation behavior occur, it is reasonable to expect that the onset temperature of a mixture lies in between, and proportional to the fraction of, the two individual components. Examining the estimated inverse mixture onset temperatures, $\beta^\mathrm{o}$, we find that they are bounded by the onset temperatures of the two individual components and change systematically depending on the fraction of each component, with the exception of sorbitol--glycerol mixtures. In the latter case we find that all the mixture data are best fit using onset temperatures that are lower than both the individual components, even as the curvature of the parabola, $J_\mathrm{m}$, is well predicted by Eq.~\ref{linear}. We revisit this pair of liquids in the discussion below.

 {\bf \em Modeling Onset Temperatures.} In the previous section we determined relaxation behavior by predicting the effective energy scale, $J_\mathrm{m}$, of the mixture. We now attempt to determine mixture relaxation times by predicting both the effective energy scale as well the onset temperature. The East model shows super-Arrhenius relaxation behavior for all $T>0$, i.e., it has no meaningful onset temperature. To model materials more realistically the temperature field is modified to $\widetilde{\beta} = \beta - \beta^\mathrm{o}$, where $\beta^\mathrm{o}$ is the inverse onset temperature of the material~\cite{keys2013calorimetric}. To more effectively predict relaxation behavior for mixtures, where each component can have a different onset temperature, we further modify our multicomponent East model so that depending on the component present each lattice site feels an effective temperature field $\widetilde{\beta}_0 = 1/T - 1/T_0^\mathrm{o}$ or $\widetilde{\beta}_1 = 1/T - 1/T_1^\mathrm{o}$, where $T_0^\mathrm{o}$ and $T_1^\mathrm{o}$ are the onset temperatures of the two components. To achieve this we modify the mixture Hamiltonian (Eq.~\ref{h}) to
\begin{equation}
H^\mathrm{o}(\{n_i,p_i\},x) = \sum_i^N n_i \Big(\frac{\widetilde{\beta}_1}{\beta} p_i J_1 + \frac{\widetilde{\beta}_0}{\beta}(1-p_i)J_0\Big) + C(\{p_i\}).
\end{equation}
The corresponding partition function, $Z^\mathrm{o}$, is given by
\begin{equation}
\begin{split}
Z^\mathrm{o} &= \sum_{n_1} \ldots \sum_{n_N} \sum_{p_1} \ldots \sum_{p_N} \exp \big\{-\sum_i^N n_i \{\widetilde{\beta}_1 J_1 p_i \\
   &\hspace{2.5cm}+ \widetilde{\beta}_0 J_0 (1-p_i) \} - \beta C(\{p_i\}) \big\},
\end{split}
\end{equation}
where it can be seen that $\widetilde{\beta}_0$ and $\widetilde{\beta}_1$ act as effective temperature fields depending on the type of species at lattice site $i$.
  Postulating again that mixtures of components behave as a single component with effective inverse onset temperature $\widetilde{\beta}^\mathrm{o}_\mathrm{m}$ and energy scale $J_\mathrm{m}$, we propose an effective single component Hamiltonian, $H_\mathrm{m}^\mathrm{o}$ of the form
\begin{equation}
H_\mathrm{m}^\mathrm{o}(\{n_i,p_i\},x) = \frac{\widetilde{\beta}_\mathrm{m}J_\mathrm{m}}{\beta}\sum_i^N n_i + C \sum_i^N p_i,
\end{equation}
where $\widetilde{\beta}_\mathrm{m}=\beta - \widetilde{\beta}^\mathrm{o}_\mathrm{m}$. The partition function, $Z_\mathrm{m}^\mathrm{o}$, is given by 
\begin{equation}
\begin{split}
Z_\mathrm{m}^\mathrm{o} &= \sum_{n_1} \ldots \sum_{n_N} \sum_{p_1} \ldots \sum_{p_N} \exp \big(-\widetilde{\beta}_\mathrm{m} J_\mathrm{m}\sum_i^N n_i \\
&\hspace{4.5cm}- \beta C \sum_i^N p_i\big),
\end{split}
\end{equation}  
where $\widetilde{\beta}_\mathrm{m}$ acts as the effective temperature field for every lattice site $i$.
 Again using Jensen's inequality to now find $\widetilde{\beta}_\mathrm{m} J_\mathrm{m}$ that best approximates $Z^\mathrm{o}$, we obtain 
\begin{equation}\label{betalinear}
\widetilde{\beta}_\mathrm{m}J_\mathrm{m} = x\widetilde{\beta}_1 J_1 + (1-x) \widetilde{\beta}_0 J_0.
\end{equation}
This equation gives an effective temperature--energy scale for a mixture given the energy scales and onset temperatures of the individual components, and reduces to our earlier result (Eq.~\ref{linear}) at low temperatures.

The effective scale $\widetilde{\beta}_\mathrm{m} J_\mathrm{m}$ can again be interpreted as giving rise to effective excitations with probability $\exp (-\widetilde{\beta}_\mathrm{m}J_\mathrm{m})$. The average distance between these excitations, i.e., the average equilibrium domain length, is $l_\mathrm{m}^\mathrm{eq} = \exp (\widetilde{\beta}_\mathrm{m} J_\mathrm{m})$, giving rise to an energy barrier of $J_\mathrm{m} \ln l^\mathrm{eq}_\mathrm{m} = \widetilde{\beta}_\mathrm{m} J_\mathrm{m}^2$. The expected time taken to overcome this barrier, and therefore the relaxation time of the average equilibrium domain length, is $\tau \sim \exp(\beta \widetilde{\beta}_\mathrm{m} J_\mathrm{m}^2).$ With reference to the relaxation time at the onset temperature, $\tau^\mathrm{o}$, we have $\tau/\tau^\mathrm{o} \sim \exp(\widetilde{\beta}_\mathrm{m}^2 J_\mathrm{m}^2)$, which is the parabolic form for relaxation of mixtures that takes into account both the energy scales, $J$, and the onset temperatures, $\beta^\mathrm{o}$, of the individual components. This form reduces to our earlier result, $\tau \sim \exp ({\beta}^2 J_\mathrm{m}^2)$, at low temperatures. 

To compare the new parabolic equation for mixtures with experimental data we use the individual values of $J$ and $\beta^\mathrm{o}$ in Eq.~\ref{betalinear} to calculate the combined energy--temperature scales. Here we only need one fitting parameter, the reference time scale $\tau^\mathrm{o}$. In Fig.~\ref{fig:allcollapsed}(c) we find that the relaxation behavior for all liquid mixtures agrees well the parabolic form when the data are scaled by $\widetilde{\beta}_\mathrm{m} J_\mathrm{m}$ as per Eq.~\ref{betalinear}, with the exception of mixtures of sorbitol and glycerol. When using Eq.~\ref{linear} in the previous section, we noted that even though the $J_\mathrm{m}$ for sorbitol--glycerol mixtures agreed well with the data, the estimated values for $\beta^\mathrm{o}$ were unexpectedly lower than $\beta^\mathrm{o}$ of the individual liquids. This, along with the lack of agreement in Fig.~\ref{fig:allcollapsed}(c), leads us to suggest further measurements for this pair of liquids. 
\newline

{\bf \em Predicting Glass Transition Temperatures of Mixtures.}  Empirical linear interpolations, such as the Gordon--Taylor equation~\cite{gordon1952ideal}, are typically used to predict the glass transition temperature for mixtures~\cite{gordon1952ideal}. This approach is not based on a microscopic physical description of glassy dynamics or glass formation, and also does not take into account the protocol dependence of the glass transition temperature. The work presented here provides a prediction for the behavior of supercooled mixtures across various temperatures, from which the glass transition temperature of a mixture can be calculated for a given cooling rate~\cite{keys2013calorimetric,hudson2018nature}. For a given cooling rate, $\nu$, the glass transition temperature, $T_\mathrm{g}$, is the temperature at which the relaxation time becomes slower than the time available to equilibrate at that temperature, i.e., the temperature at which $\nu^{-1} \sim |{d\tau}/{dT}|_{T=T_\mathrm{g}}$.
%
Using the parabolic expression $\tau\sim\tau^\mathrm{o}\exp(\widetilde{\beta}_\mathrm{m}^2 J_\mathrm{m}^2)$ in the above equation gives a prediction for the glass transition temperature of a mixture. 

In summary, we find that a mixture of well-mixed glass-forming liquids behaves as a single material whose dynamics is governed by an effective energy scale, an average of the energy scales of the individual components. We anticipate the use of our effective model in predicting transport properties and glass transition temperatures of mixtures especially for compounds, and under conditions, where making measurements is challenging. Extending this model from binary to polydisperse mixtures may provide insight into the effects of polydispersity on the dynamics of glass-formers~\cite{berthier2016equilibrium}. We also anticipate the use of this model in studying the interplay between dynamical heterogeneity of the system and the diffusion and phase separation of individual components, possibly involving interfacial fronts between two dynamical phases bearing a dynamical interfacial tension~\cite{katira2018solvation}. This is especially relevant in multi-component systems such as atmospheric organic aerosols, where melting (and formation) of amorphous solids is hypothesized to occur via an increase (or decrease) in water content~\cite{koop2011glass}.

\begin{acknowledgments}
We thank David Chandler for introducing us to this problem and for discussions during the early part of this work. S.K. acknowledges funding from the National Science Foundation (award number 1507642) and the University of California, Berkeley. K.K.M acknowledges funding from the Department of Energy (contract DE-AC02-05CH11231, FWP no. CHPHYS02). J.P.G was supported by EPSRC Grant no.\ EP/R04421X/1.
\end{acknowledgments}


\begin{thebibliography}{40}%
\makeatletter
\providecommand \@ifxundefined [1]{%
 \@ifx{#1\undefined}
}%
\providecommand \@ifnum [1]{%
 \ifnum #1\expandafter \@firstoftwo
 \else \expandafter \@secondoftwo
 \fi
}%
\providecommand \@ifx [1]{%
 \ifx #1\expandafter \@firstoftwo
 \else \expandafter \@secondoftwo
 \fi
}%
\providecommand \natexlab [1]{#1}%
\providecommand \enquote  [1]{``#1''}%
\providecommand \bibnamefont  [1]{#1}%
\providecommand \bibfnamefont [1]{#1}%
\providecommand \citenamefont [1]{#1}%
\providecommand \href@noop [0]{\@secondoftwo}%
\providecommand \href [0]{\begingroup \@sanitize@url \@href}%
\providecommand \@href[1]{\@@startlink{#1}\@@href}%
\providecommand \@@href[1]{\endgroup#1\@@endlink}%
\providecommand \@sanitize@url [0]{\catcode `\\12\catcode `\$12\catcode
  `\&12\catcode `\#12\catcode `\^12\catcode `\_12\catcode `\%12\relax}%
\providecommand \@@startlink[1]{}%
\providecommand \@@endlink[0]{}%
\providecommand \url  [0]{\begingroup\@sanitize@url \@url }%
\providecommand \@url [1]{\endgroup\@href {#1}{\urlprefix }}%
\providecommand \urlprefix  [0]{URL }%
\providecommand \Eprint [0]{\href }%
\providecommand \doibase [0]{http://dx.doi.org/}%
\providecommand \selectlanguage [0]{\@gobble}%
\providecommand \bibinfo  [0]{\@secondoftwo}%
\providecommand \bibfield  [0]{\@secondoftwo}%
\providecommand \translation [1]{[#1]}%
\providecommand \BibitemOpen [0]{}%
\providecommand \bibitemStop [0]{}%
\providecommand \bibitemNoStop [0]{.\EOS\space}%
\providecommand \EOS [0]{\spacefactor3000\relax}%
\providecommand \BibitemShut  [1]{\csname bibitem#1\endcsname}%
\let\auto@bib@innerbib\@empty
\bibitem [{\citenamefont {Blochowicz}\ and\ \citenamefont
  {R{\"o}ssler}(2004)}]{blochowicz2004beta}%
  \BibitemOpen
  \bibfield  {author} {\bibinfo {author} {\bibfnamefont {T.}~\bibnamefont
  {Blochowicz}}\ and\ \bibinfo {author} {\bibfnamefont {E.}~\bibnamefont
  {R{\"o}ssler}},\ }\href@noop {} {\bibfield  {journal} {\bibinfo  {journal}
  {Physical Review Letters}\ }\textbf {\bibinfo {volume} {92}},\ \bibinfo
  {pages} {225701} (\bibinfo {year} {2004})}\BibitemShut {NoStop}%
\bibitem [{\citenamefont {Wang}\ \emph {et~al.}(2010)\citenamefont {Wang},
  \citenamefont {Tian}, \citenamefont {Liu},\ and\ \citenamefont
  {Richert}}]{wang2010structural}%
  \BibitemOpen
  \bibfield  {author} {\bibinfo {author} {\bibfnamefont {L.-M.}\ \bibnamefont
  {Wang}}, \bibinfo {author} {\bibfnamefont {Y.}~\bibnamefont {Tian}}, \bibinfo
  {author} {\bibfnamefont {R.}~\bibnamefont {Liu}}, \ and\ \bibinfo {author}
  {\bibfnamefont {R.}~\bibnamefont {Richert}},\ }\href@noop {} {\bibfield
  {journal} {\bibinfo  {journal} {The Journal of Physical Chemistry B}\
  }\textbf {\bibinfo {volume} {114}},\ \bibinfo {pages} {3618} (\bibinfo {year}
  {2010})}\BibitemShut {NoStop}%
\bibitem [{\citenamefont {Foffi}\ \emph {et~al.}(2003)\citenamefont {Foffi},
  \citenamefont {G{\"o}tze}, \citenamefont {Sciortino}, \citenamefont
  {Tartaglia},\ and\ \citenamefont {Voigtmann}}]{foffi2003mixing}%
  \BibitemOpen
  \bibfield  {author} {\bibinfo {author} {\bibfnamefont {G.}~\bibnamefont
  {Foffi}}, \bibinfo {author} {\bibfnamefont {W.}~\bibnamefont {G{\"o}tze}},
  \bibinfo {author} {\bibfnamefont {F.}~\bibnamefont {Sciortino}}, \bibinfo
  {author} {\bibfnamefont {P.}~\bibnamefont {Tartaglia}}, \ and\ \bibinfo
  {author} {\bibfnamefont {T.}~\bibnamefont {Voigtmann}},\ }\href@noop {}
  {\bibfield  {journal} {\bibinfo  {journal} {Physical Review Letters}\
  }\textbf {\bibinfo {volume} {91}},\ \bibinfo {pages} {085701} (\bibinfo
  {year} {2003})}\BibitemShut {NoStop}%
\bibitem [{\citenamefont {Angell}\ \emph {et~al.}(1978)\citenamefont {Angell},
  \citenamefont {Sare},\ and\ \citenamefont {Sare}}]{angell1978glass}%
  \BibitemOpen
  \bibfield  {author} {\bibinfo {author} {\bibfnamefont {C.}~\bibnamefont
  {Angell}}, \bibinfo {author} {\bibfnamefont {J.}~\bibnamefont {Sare}}, \ and\
  \bibinfo {author} {\bibfnamefont {E.}~\bibnamefont {Sare}},\ }\href@noop {}
  {\bibfield  {journal} {\bibinfo  {journal} {The Journal of Physical
  Chemistry}\ }\textbf {\bibinfo {volume} {82}},\ \bibinfo {pages} {2622}
  (\bibinfo {year} {1978})}\BibitemShut {NoStop}%
\bibitem [{\citenamefont {Hancock}\ and\ \citenamefont
  {Zografi}(1997)}]{hancock1997characteristics}%
  \BibitemOpen
  \bibfield  {author} {\bibinfo {author} {\bibfnamefont {B.~C.}\ \bibnamefont
  {Hancock}}\ and\ \bibinfo {author} {\bibfnamefont {G.}~\bibnamefont
  {Zografi}},\ }\href@noop {} {\bibfield  {journal} {\bibinfo  {journal}
  {Journal of Pharmaceutical Sciences}\ }\textbf {\bibinfo {volume} {86}},\
  \bibinfo {pages} {1} (\bibinfo {year} {1997})}\BibitemShut {NoStop}%
\bibitem [{\citenamefont {Yu}(2001)}]{yu2001amorphous}%
  \BibitemOpen
  \bibfield  {author} {\bibinfo {author} {\bibfnamefont {L.}~\bibnamefont
  {Yu}},\ }\href@noop {} {\bibfield  {journal} {\bibinfo  {journal} {Advanced
  Drug Delivery Reviews}\ }\textbf {\bibinfo {volume} {48}},\ \bibinfo {pages}
  {27} (\bibinfo {year} {2001})}\BibitemShut {NoStop}%
\bibitem [{\citenamefont {Bhandari}\ and\ \citenamefont
  {Howes}(1999)}]{bhandari1999implication}%
  \BibitemOpen
  \bibfield  {author} {\bibinfo {author} {\bibfnamefont {B.}~\bibnamefont
  {Bhandari}}\ and\ \bibinfo {author} {\bibfnamefont {T.}~\bibnamefont
  {Howes}},\ }\href@noop {} {\bibfield  {journal} {\bibinfo  {journal} {Journal
  of Food Engineering}\ }\textbf {\bibinfo {volume} {40}},\ \bibinfo {pages}
  {71} (\bibinfo {year} {1999})}\BibitemShut {NoStop}%
\bibitem [{\citenamefont {Virtanen}\ \emph {et~al.}(2010)\citenamefont
  {Virtanen}, \citenamefont {Joutsensaari}, \citenamefont {Koop}, \citenamefont
  {Kannosto}, \citenamefont {Yli-Piril{\"a}}, \citenamefont {Leskinen},
  \citenamefont {M{\"a}kel{\"a}}, \citenamefont {Holopainen}, \citenamefont
  {P{\"o}schl}, \citenamefont {Kulmala} \emph
  {et~al.}}]{virtanen2010amorphous}%
  \BibitemOpen
  \bibfield  {author} {\bibinfo {author} {\bibfnamefont {A.}~\bibnamefont
  {Virtanen}}, \bibinfo {author} {\bibfnamefont {J.}~\bibnamefont
  {Joutsensaari}}, \bibinfo {author} {\bibfnamefont {T.}~\bibnamefont {Koop}},
  \bibinfo {author} {\bibfnamefont {J.}~\bibnamefont {Kannosto}}, \bibinfo
  {author} {\bibfnamefont {P.}~\bibnamefont {Yli-Piril{\"a}}}, \bibinfo
  {author} {\bibfnamefont {J.}~\bibnamefont {Leskinen}}, \bibinfo {author}
  {\bibfnamefont {J.~M.}\ \bibnamefont {M{\"a}kel{\"a}}}, \bibinfo {author}
  {\bibfnamefont {J.~K.}\ \bibnamefont {Holopainen}}, \bibinfo {author}
  {\bibfnamefont {U.}~\bibnamefont {P{\"o}schl}}, \bibinfo {author}
  {\bibfnamefont {M.}~\bibnamefont {Kulmala}},  \emph {et~al.},\ }\href@noop {}
  {\bibfield  {journal} {\bibinfo  {journal} {Nature}\ }\textbf {\bibinfo
  {volume} {467}},\ \bibinfo {pages} {824} (\bibinfo {year}
  {2010})}\BibitemShut {NoStop}%
\bibitem [{\citenamefont {Vaden}\ \emph {et~al.}(2010)\citenamefont {Vaden},
  \citenamefont {Song}, \citenamefont {Zaveri}, \citenamefont {Imre},\ and\
  \citenamefont {Zelenyuk}}]{vaden2010morphology}%
  \BibitemOpen
  \bibfield  {author} {\bibinfo {author} {\bibfnamefont {T.~D.}\ \bibnamefont
  {Vaden}}, \bibinfo {author} {\bibfnamefont {C.}~\bibnamefont {Song}},
  \bibinfo {author} {\bibfnamefont {R.~A.}\ \bibnamefont {Zaveri}}, \bibinfo
  {author} {\bibfnamefont {D.}~\bibnamefont {Imre}}, \ and\ \bibinfo {author}
  {\bibfnamefont {A.}~\bibnamefont {Zelenyuk}},\ }\href@noop {} {\bibfield
  {journal} {\bibinfo  {journal} {Proceedings of the National Academy of
  Sciences}\ }\textbf {\bibinfo {volume} {107}},\ \bibinfo {pages} {6658}
  (\bibinfo {year} {2010})}\BibitemShut {NoStop}%
\bibitem [{\citenamefont {Cappa}\ and\ \citenamefont
  {Wilson}(2011)}]{cappa2011evolution}%
  \BibitemOpen
  \bibfield  {author} {\bibinfo {author} {\bibfnamefont {C.~D.}\ \bibnamefont
  {Cappa}}\ and\ \bibinfo {author} {\bibfnamefont {K.~R.}\ \bibnamefont
  {Wilson}},\ }\href@noop {} {\bibfield  {journal} {\bibinfo  {journal}
  {Atmospheric Chemistry and Physics}\ }\textbf {\bibinfo {volume} {11}},\
  \bibinfo {pages} {1895} (\bibinfo {year} {2011})}\BibitemShut {NoStop}%
\bibitem [{\citenamefont {Koop}\ \emph {et~al.}(2011)\citenamefont {Koop},
  \citenamefont {Bookhold}, \citenamefont {Shiraiwa},\ and\ \citenamefont
  {P{\"o}schl}}]{koop2011glass}%
  \BibitemOpen
  \bibfield  {author} {\bibinfo {author} {\bibfnamefont {T.}~\bibnamefont
  {Koop}}, \bibinfo {author} {\bibfnamefont {J.}~\bibnamefont {Bookhold}},
  \bibinfo {author} {\bibfnamefont {M.}~\bibnamefont {Shiraiwa}}, \ and\
  \bibinfo {author} {\bibfnamefont {U.}~\bibnamefont {P{\"o}schl}},\
  }\href@noop {} {\bibfield  {journal} {\bibinfo  {journal} {Physical Chemistry
  Chemical Physics}\ }\textbf {\bibinfo {volume} {13}},\ \bibinfo {pages}
  {19238} (\bibinfo {year} {2011})}\BibitemShut {NoStop}%
\bibitem [{\citenamefont {Crowe}\ \emph {et~al.}(1998)\citenamefont {Crowe},
  \citenamefont {Carpenter},\ and\ \citenamefont {Crowe}}]{crowe1998role}%
  \BibitemOpen
  \bibfield  {author} {\bibinfo {author} {\bibfnamefont {J.~H.}\ \bibnamefont
  {Crowe}}, \bibinfo {author} {\bibfnamefont {J.~F.}\ \bibnamefont
  {Carpenter}}, \ and\ \bibinfo {author} {\bibfnamefont {L.~M.}\ \bibnamefont
  {Crowe}},\ }\href@noop {} {\bibfield  {journal} {\bibinfo  {journal} {Annual
  Review of Physiology}\ }\textbf {\bibinfo {volume} {60}},\ \bibinfo {pages}
  {73} (\bibinfo {year} {1998})}\BibitemShut {NoStop}%
\bibitem [{\citenamefont {Green}\ and\ \citenamefont
  {Angell}(1989)}]{green1989phase}%
  \BibitemOpen
  \bibfield  {author} {\bibinfo {author} {\bibfnamefont {J.~L.}\ \bibnamefont
  {Green}}\ and\ \bibinfo {author} {\bibfnamefont {C.~A.}\ \bibnamefont
  {Angell}},\ }\href@noop {} {\bibfield  {journal} {\bibinfo  {journal} {The
  Journal of Physical Chemistry}\ }\textbf {\bibinfo {volume} {93}},\ \bibinfo
  {pages} {2880} (\bibinfo {year} {1989})}\BibitemShut {NoStop}%
\bibitem [{\citenamefont {Gordon}\ and\ \citenamefont
  {Taylor}(1952)}]{gordon1952ideal}%
  \BibitemOpen
  \bibfield  {author} {\bibinfo {author} {\bibfnamefont {M.}~\bibnamefont
  {Gordon}}\ and\ \bibinfo {author} {\bibfnamefont {J.~S.}\ \bibnamefont
  {Taylor}},\ }\href@noop {} {\bibfield  {journal} {\bibinfo  {journal}
  {Journal of Applied Chemistry}\ }\textbf {\bibinfo {volume} {2}},\ \bibinfo
  {pages} {493} (\bibinfo {year} {1952})}\BibitemShut {NoStop}%
\bibitem [{\citenamefont {Ediger}\ \emph {et~al.}(1996)\citenamefont {Ediger},
  \citenamefont {Angell},\ and\ \citenamefont {Nagel}}]{ediger1996supercooled}%
  \BibitemOpen
  \bibfield  {author} {\bibinfo {author} {\bibfnamefont {M.~D.}\ \bibnamefont
  {Ediger}}, \bibinfo {author} {\bibfnamefont {C.~A.}\ \bibnamefont {Angell}},
  \ and\ \bibinfo {author} {\bibfnamefont {S.~R.}\ \bibnamefont {Nagel}},\
  }\href@noop {} {\bibfield  {journal} {\bibinfo  {journal} {The Journal of
  Physical Chemistry}\ }\textbf {\bibinfo {volume} {100}},\ \bibinfo {pages}
  {13200} (\bibinfo {year} {1996})}\BibitemShut {NoStop}%
\bibitem [{\citenamefont {Angell}\ \emph {et~al.}(2000)\citenamefont {Angell},
  \citenamefont {Ngai}, \citenamefont {McKenna}, \citenamefont {McMillan},\
  and\ \citenamefont {Martin}}]{angell2000relaxation}%
  \BibitemOpen
  \bibfield  {author} {\bibinfo {author} {\bibfnamefont {C.~A.}\ \bibnamefont
  {Angell}}, \bibinfo {author} {\bibfnamefont {K.~L.}\ \bibnamefont {Ngai}},
  \bibinfo {author} {\bibfnamefont {G.~B.}\ \bibnamefont {McKenna}}, \bibinfo
  {author} {\bibfnamefont {P.~F.}\ \bibnamefont {McMillan}}, \ and\ \bibinfo
  {author} {\bibfnamefont {S.~W.}\ \bibnamefont {Martin}},\ }\href@noop {}
  {\bibfield  {journal} {\bibinfo  {journal} {Journal of Applied Physics}\
  }\textbf {\bibinfo {volume} {88}},\ \bibinfo {pages} {3113} (\bibinfo {year}
  {2000})}\BibitemShut {NoStop}%
\bibitem [{\citenamefont {Palmer}\ \emph {et~al.}(1984)\citenamefont {Palmer},
  \citenamefont {Stein}, \citenamefont {Abrahams},\ and\ \citenamefont
  {Anderson}}]{palmer1984models}%
  \BibitemOpen
  \bibfield  {author} {\bibinfo {author} {\bibfnamefont {R.~G.}\ \bibnamefont
  {Palmer}}, \bibinfo {author} {\bibfnamefont {D.~L.}\ \bibnamefont {Stein}},
  \bibinfo {author} {\bibfnamefont {E.}~\bibnamefont {Abrahams}}, \ and\
  \bibinfo {author} {\bibfnamefont {P.~W.}\ \bibnamefont {Anderson}},\
  }\href@noop {} {\bibfield  {journal} {\bibinfo  {journal} {Physical Review
  Letters}\ }\textbf {\bibinfo {volume} {53}},\ \bibinfo {pages} {958}
  (\bibinfo {year} {1984})}\BibitemShut {NoStop}%
\bibitem [{\citenamefont {Sollich}\ and\ \citenamefont
  {Evans}(1999)}]{sollich1999glassy}%
  \BibitemOpen
  \bibfield  {author} {\bibinfo {author} {\bibfnamefont {P.}~\bibnamefont
  {Sollich}}\ and\ \bibinfo {author} {\bibfnamefont {M.~R.}\ \bibnamefont
  {Evans}},\ }\href@noop {} {\bibfield  {journal} {\bibinfo  {journal}
  {Physical Review Letters}\ }\textbf {\bibinfo {volume} {83}},\ \bibinfo
  {pages} {3238} (\bibinfo {year} {1999})}\BibitemShut {NoStop}%
\bibitem [{\citenamefont {Sollich}\ and\ \citenamefont
  {Evans}(2003)}]{sollich2003glassy}%
  \BibitemOpen
  \bibfield  {author} {\bibinfo {author} {\bibfnamefont {P.}~\bibnamefont
  {Sollich}}\ and\ \bibinfo {author} {\bibfnamefont {M.}~\bibnamefont
  {Evans}},\ }\href@noop {} {\bibfield  {journal} {\bibinfo  {journal}
  {Physical Review E}\ }\textbf {\bibinfo {volume} {68}},\ \bibinfo {pages}
  {031504} (\bibinfo {year} {2003})}\BibitemShut {NoStop}%
\bibitem [{\citenamefont {Garrahan}\ and\ \citenamefont
  {Chandler}(2002)}]{garrahan2002geometrical}%
  \BibitemOpen
  \bibfield  {author} {\bibinfo {author} {\bibfnamefont {J.~P.}\ \bibnamefont
  {Garrahan}}\ and\ \bibinfo {author} {\bibfnamefont {D.}~\bibnamefont
  {Chandler}},\ }\href@noop {} {\bibfield  {journal} {\bibinfo  {journal}
  {Physical Review Letters}\ }\textbf {\bibinfo {volume} {89}},\ \bibinfo
  {pages} {035704} (\bibinfo {year} {2002})}\BibitemShut {NoStop}%
\bibitem [{\citenamefont {Chandler}\ and\ \citenamefont
  {Garrahan}(2010)}]{chandler2010dynamics}%
  \BibitemOpen
  \bibfield  {author} {\bibinfo {author} {\bibfnamefont {D.}~\bibnamefont
  {Chandler}}\ and\ \bibinfo {author} {\bibfnamefont {J.~P.}\ \bibnamefont
  {Garrahan}},\ }\href@noop {} {\bibfield  {journal} {\bibinfo  {journal}
  {Annual Review of Physical Chemistry}\ }\textbf {\bibinfo {volume} {61}},\
  \bibinfo {pages} {191} (\bibinfo {year} {2010})}\BibitemShut {NoStop}%
\bibitem [{\citenamefont {Keys}\ \emph {et~al.}(2011)\citenamefont {Keys},
  \citenamefont {Hedges}, \citenamefont {Garrahan}, \citenamefont {Glotzer},\
  and\ \citenamefont {Chandler}}]{keys2011}%
  \BibitemOpen
  \bibfield  {author} {\bibinfo {author} {\bibfnamefont {A.~S.}\ \bibnamefont
  {Keys}}, \bibinfo {author} {\bibfnamefont {L.~O.}\ \bibnamefont {Hedges}},
  \bibinfo {author} {\bibfnamefont {J.~P.}\ \bibnamefont {Garrahan}}, \bibinfo
  {author} {\bibfnamefont {S.~C.}\ \bibnamefont {Glotzer}}, \ and\ \bibinfo
  {author} {\bibfnamefont {D.}~\bibnamefont {Chandler}},\ }\href@noop {}
  {\bibfield  {journal} {\bibinfo  {journal} {Physical Review X}\ }\textbf
  {\bibinfo {volume} {1}},\ \bibinfo {pages} {021013} (\bibinfo {year}
  {2011})}\BibitemShut {NoStop}%
  \bibitem [{\citenamefont {Speck}(2019)}]{speck2019dynamic}%
  \BibitemOpen
  \bibfield  {author} {\bibinfo {author} {\bibfnamefont {T.}~\bibnamefont
  {Speck}},\ }\href@noop {} {\bibfield  {journal} {\bibinfo  {journal} {arXiv
  preprint arXiv:1902.07768}\ } (\bibinfo {year} {2019})}\BibitemShut {NoStop}%
\bibitem [{\citenamefont {Elmatad}\ \emph {et~al.}(2009)\citenamefont
  {Elmatad}, \citenamefont {Chandler},\ and\ \citenamefont
  {Garrahan}}]{elmatad2009corresponding}%
  \BibitemOpen
  \bibfield  {author} {\bibinfo {author} {\bibfnamefont {Y.~S.}\ \bibnamefont
  {Elmatad}}, \bibinfo {author} {\bibfnamefont {D.}~\bibnamefont {Chandler}}, \
  and\ \bibinfo {author} {\bibfnamefont {J.~P.}\ \bibnamefont {Garrahan}},\
  }\href@noop {} {\bibfield  {journal} {\bibinfo  {journal} {The Journal of
  Physical Chemistry B}\ }\textbf {\bibinfo {volume} {113}},\ \bibinfo {pages}
  {5563} (\bibinfo {year} {2009})}\BibitemShut {NoStop}%
\bibitem [{\citenamefont {Keys}\ \emph {et~al.}(2013)\citenamefont {Keys},
  \citenamefont {Garrahan},\ and\ \citenamefont
  {Chandler}}]{keys2013calorimetric}%
  \BibitemOpen
  \bibfield  {author} {\bibinfo {author} {\bibfnamefont {A.~S.}\ \bibnamefont
  {Keys}}, \bibinfo {author} {\bibfnamefont {J.~P.}\ \bibnamefont {Garrahan}},
  \ and\ \bibinfo {author} {\bibfnamefont {D.}~\bibnamefont {Chandler}},\
  }\href@noop {} {\bibfield  {journal} {\bibinfo  {journal} {Proceedings of the
  National Academy of Sciences}\ ,\ \bibinfo {pages} {201302665}} (\bibinfo
  {year} {2013})}\BibitemShut {NoStop}%
\bibitem [{\citenamefont {Hudson}\ and\ \citenamefont
  {Mandadapu}(2018)}]{hudson2018nature}%
  \BibitemOpen
  \bibfield  {author} {\bibinfo {author} {\bibfnamefont {A.}~\bibnamefont
  {Hudson}}\ and\ \bibinfo {author} {\bibfnamefont {K.~K.}\ \bibnamefont
  {Mandadapu}},\ }\href@noop {} {\bibfield  {journal} {\bibinfo  {journal}
  {arXiv preprint arXiv:1804.03769}\ } (\bibinfo {year} {2018})}\BibitemShut
  {NoStop}%
\bibitem [{\citenamefont {Adam}\ and\ \citenamefont
  {Gibbs}(1965)}]{adam1965temperature}%
  \BibitemOpen
  \bibfield  {author} {\bibinfo {author} {\bibfnamefont {G.}~\bibnamefont
  {Adam}}\ and\ \bibinfo {author} {\bibfnamefont {J.~H.}\ \bibnamefont
  {Gibbs}},\ }\href@noop {} {\bibfield  {journal} {\bibinfo  {journal} {The
  Journal of Chemical Physics}\ }\textbf {\bibinfo {volume} {43}},\ \bibinfo
  {pages} {139} (\bibinfo {year} {1965})}\BibitemShut {NoStop}%
\bibitem [{\citenamefont {Gotze}\ and\ \citenamefont
  {Sjogren}(1992)}]{gotze1992relaxation}%
  \BibitemOpen
  \bibfield  {author} {\bibinfo {author} {\bibfnamefont {W.}~\bibnamefont
  {Gotze}}\ and\ \bibinfo {author} {\bibfnamefont {L.}~\bibnamefont
  {Sjogren}},\ }\href@noop {} {\bibfield  {journal} {\bibinfo  {journal}
  {Reports on Progress in Physics}\ }\textbf {\bibinfo {volume} {55}},\
  \bibinfo {pages} {241} (\bibinfo {year} {1992})}\BibitemShut {NoStop}%
\bibitem [{\citenamefont {M{\'e}zard}\ and\ \citenamefont
  {Parisi}(1999)}]{mezard1999thermodynamics}%
  \BibitemOpen
  \bibfield  {author} {\bibinfo {author} {\bibfnamefont {M.}~\bibnamefont
  {M{\'e}zard}}\ and\ \bibinfo {author} {\bibfnamefont {G.}~\bibnamefont
  {Parisi}},\ }\href@noop {} {\bibfield  {journal} {\bibinfo  {journal}
  {Journal of Physics: Condensed Matter}\ }\textbf {\bibinfo {volume} {11}},\
  \bibinfo {pages} {A157} (\bibinfo {year} {1999})}\BibitemShut {NoStop}%
\bibitem [{\citenamefont {Xia}\ and\ \citenamefont
  {Wolynes}(2001)}]{xia2001microscopic}%
  \BibitemOpen
  \bibfield  {author} {\bibinfo {author} {\bibfnamefont {X.}~\bibnamefont
  {Xia}}\ and\ \bibinfo {author} {\bibfnamefont {P.~G.}\ \bibnamefont
  {Wolynes}},\ }\href@noop {} {\bibfield  {journal} {\bibinfo  {journal}
  {Physical Review Letters}\ }\textbf {\bibinfo {volume} {86}},\ \bibinfo
  {pages} {5526} (\bibinfo {year} {2001})}\BibitemShut {NoStop}%
  \bibitem [{\citenamefont {Ritort}\ and\ \citenamefont
  {Sollich}(2003)}]{Ritort2003}%
  \BibitemOpen
  \bibfield  {author} {\bibinfo {author} {\bibfnamefont {F.}~\bibnamefont
  {Ritort}}\ and\ \bibinfo {author} {\bibfnamefont {P.}~\bibnamefont
  {Sollich}},\ }\href@noop {} {\bibfield  {journal} {\bibinfo  {journal} {Adv.
  Phys.}\ }\textbf {\bibinfo {volume} {52}},\ \bibinfo {pages} {219} (\bibinfo
  {year} {2003})}\BibitemShut {NoStop}%
\bibitem [{\citenamefont {Garrahan}\ \emph {et~al.}(2011)\citenamefont
  {Garrahan}, \citenamefont {Sollich},\ and\ \citenamefont
  {Toninelli}}]{Garrahan2011}%
  \BibitemOpen
  \bibfield  {author} {\bibinfo {author} {\bibfnamefont {J.~P.}\ \bibnamefont
  {Garrahan}}, \bibinfo {author} {\bibfnamefont {P.}~\bibnamefont {Sollich}}, \
  and\ \bibinfo {author} {\bibfnamefont {C.}~\bibnamefont {Toninelli}},\ }in\
  \href@noop {} {\emph {\bibinfo {booktitle} {Dynamical Heterogeneities in
  Glasses, Colloids, and Granular Media}}},\ \bibinfo {series and number}
  {International Series of Monographs on Physics},\ \bibinfo {editor} {edited
  by\ \bibinfo {editor} {\bibfnamefont {L.}~\bibnamefont {Berthier}}, \bibinfo
  {editor} {\bibfnamefont {G.}~\bibnamefont {Biroli}}, \bibinfo {editor}
  {\bibfnamefont {J.-P.}\ \bibnamefont {Bouchaud}}, \bibinfo {editor}
  {\bibfnamefont {L.}~\bibnamefont {Cipelletti}}, \ and\ \bibinfo {editor}
  {\bibfnamefont {W.}~\bibnamefont {van Saarloos}}}\ (\bibinfo  {publisher}
  {Oxford University Press},\ \bibinfo {address} {Oxford, UK},\ \bibinfo {year}
  {2011})\BibitemShut {NoStop}%
\bibitem [{\citenamefont {Garrahan}(2018)}]{Garrahan2018}%
  \BibitemOpen
  \bibfield  {author} {\bibinfo {author} {\bibfnamefont {J.~P.}\ \bibnamefont
  {Garrahan}},\ }\href {\doibase 10.1016/j.physa.2017.12.149} {\bibfield
  {journal} {\bibinfo  {journal} {Physica A}\ }\textbf {\bibinfo {volume}
  {504}},\ \bibinfo {pages} {130} (\bibinfo {year} {2018})}\BibitemShut
  {NoStop}%

\bibitem [{\citenamefont {J{\"a}ckle}\ and\ \citenamefont
  {Eisinger}(1991)}]{jackle1991hierarchically}%
  \BibitemOpen
  \bibfield  {author} {\bibinfo {author} {\bibfnamefont {J.}~\bibnamefont
  {J{\"a}ckle}}\ and\ \bibinfo {author} {\bibfnamefont {S.}~\bibnamefont
  {Eisinger}},\ }\href@noop {} {\bibfield  {journal} {\bibinfo  {journal}
  {Zeitschrift f{\"u}r Physik B Condensed Matter}\ }\textbf {\bibinfo {volume}
  {84}},\ \bibinfo {pages} {115} (\bibinfo {year} {1991})}\BibitemShut
  {NoStop}%
\bibitem [{\citenamefont {Chleboun}\ \emph {et~al.}(2013)\citenamefont
  {Chleboun}, \citenamefont {Faggionato},\ and\ \citenamefont
  {Martinelli}}]{chleboun2013time}%
  \BibitemOpen
  \bibfield  {author} {\bibinfo {author} {\bibfnamefont {P.}~\bibnamefont
  {Chleboun}}, \bibinfo {author} {\bibfnamefont {A.}~\bibnamefont
  {Faggionato}}, \ and\ \bibinfo {author} {\bibfnamefont {F.}~\bibnamefont
  {Martinelli}},\ }\href@noop {} {\bibfield  {journal} {\bibinfo  {journal}
  {Journal of Statistical Mechanics: Theory and Experiment}\ }\textbf {\bibinfo
  {volume} {2013}},\ \bibinfo {pages} {L04001} (\bibinfo {year}
  {2013})}\BibitemShut {NoStop}%
\bibitem [{\citenamefont {Feynman}(2018)}]{feynman2018statistical}%
  \BibitemOpen
  \bibfield  {author} {\bibinfo {author} {\bibfnamefont {R.~P.}\ \bibnamefont
  {Feynman}},\ }\href@noop {} {\emph {\bibinfo {title} {Statistical Mechanics:
  A Set of Lectures}}}\ (\bibinfo  {publisher} {CRC Press},\ \bibinfo {year}
  {2018})\BibitemShut {NoStop}%
\bibitem [{\citenamefont {Chandler}(1987)}]{chandler1987introduction}%
  \BibitemOpen
  \bibfield  {author} {\bibinfo {author} {\bibfnamefont {D.}~\bibnamefont
  {Chandler}},\ }\href@noop {} {\emph {\bibinfo {title} {Introduction to Modern
  Statistical Mechanics}}}\ (\bibinfo {year} {1987})\BibitemShut {NoStop}%
\bibitem [{\citenamefont {Duvvuri}\ and\ \citenamefont
  {Richert}(2004)}]{duvvuri2004binary}%
  \BibitemOpen
  \bibfield  {author} {\bibinfo {author} {\bibfnamefont {K.}~\bibnamefont
  {Duvvuri}}\ and\ \bibinfo {author} {\bibfnamefont {R.}~\bibnamefont
  {Richert}},\ }\href@noop {} {\bibfield  {journal} {\bibinfo  {journal} {The
  Journal of Physical Chemistry B}\ }\textbf {\bibinfo {volume} {108}},\
  \bibinfo {pages} {10451} (\bibinfo {year} {2004})}\BibitemShut {NoStop}%
\bibitem [{\citenamefont {Wang}\ and\ \citenamefont
  {Richert}(2005)}]{wang2005ideal}%
  \BibitemOpen
  \bibfield  {author} {\bibinfo {author} {\bibfnamefont {L.-M.}\ \bibnamefont
  {Wang}}\ and\ \bibinfo {author} {\bibfnamefont {R.}~\bibnamefont {Richert}},\
  }\href@noop {} {\bibfield  {journal} {\bibinfo  {journal} {The Journal of
  Physical Chemistry B}\ }\textbf {\bibinfo {volume} {109}},\ \bibinfo {pages}
  {8767} (\bibinfo {year} {2005})}\BibitemShut {NoStop}%
\bibitem [{\citenamefont {Puzenko}\ \emph {et~al.}(2005)\citenamefont
  {Puzenko}, \citenamefont {Hayashi}, \citenamefont {Ryabov}, \citenamefont
  {Balin}, \citenamefont {Feldman}, \citenamefont {Kaatze},\ and\ \citenamefont
  {Behrends}}]{puzenko2005relaxation}%
  \BibitemOpen
  \bibfield  {author} {\bibinfo {author} {\bibfnamefont {A.}~\bibnamefont
  {Puzenko}}, \bibinfo {author} {\bibfnamefont {Y.}~\bibnamefont {Hayashi}},
  \bibinfo {author} {\bibfnamefont {Y.~E.}\ \bibnamefont {Ryabov}}, \bibinfo
  {author} {\bibfnamefont {I.}~\bibnamefont {Balin}}, \bibinfo {author}
  {\bibfnamefont {Y.}~\bibnamefont {Feldman}}, \bibinfo {author} {\bibfnamefont
  {U.}~\bibnamefont {Kaatze}}, \ and\ \bibinfo {author} {\bibfnamefont
  {R.}~\bibnamefont {Behrends}},\ }\href@noop {} {\bibfield  {journal}
  {\bibinfo  {journal} {The Journal of Physical Chemistry B}\ }\textbf
  {\bibinfo {volume} {109}},\ \bibinfo {pages} {6031} (\bibinfo {year}
  {2005})}\BibitemShut {NoStop}%
\bibitem [{\citenamefont {Schuster}(2016)}]{schuster2016glassy}%
  \BibitemOpen
  \bibfield  {author} {\bibinfo {author} {\bibfnamefont {K.~C.}\ \bibnamefont
  {Schuster}},\ }\emph {\bibinfo {title} {Glassy Dynamics on a Lattice and in
  Nature}},\ \href@noop {} {Ph.D. thesis},\ \bibinfo  {school} {UC Berkeley}
  (\bibinfo {year} {2016})\BibitemShut {NoStop}%
\bibitem [{\citenamefont {Wagner}\ and\ \citenamefont
  {Richert}(1999)}]{wagner1999equilibrium}%
  \BibitemOpen
  \bibfield  {author} {\bibinfo {author} {\bibfnamefont {H.}~\bibnamefont
  {Wagner}}\ and\ \bibinfo {author} {\bibfnamefont {R.}~\bibnamefont
  {Richert}},\ }\href@noop {} {\bibfield  {journal} {\bibinfo  {journal} {The
  Journal of Physical Chemistry B}\ }\textbf {\bibinfo {volume} {103}},\
  \bibinfo {pages} {4071} (\bibinfo {year} {1999})}\BibitemShut {NoStop}%
\bibitem [{\citenamefont {Berthier}\ \emph {et~al.}(2016)\citenamefont
  {Berthier}, \citenamefont {Coslovich}, \citenamefont {Ninarello},\ and\
  \citenamefont {Ozawa}}]{berthier2016equilibrium}%
  \BibitemOpen
  \bibfield  {author} {\bibinfo {author} {\bibfnamefont {L.}~\bibnamefont
  {Berthier}}, \bibinfo {author} {\bibfnamefont {D.}~\bibnamefont {Coslovich}},
  \bibinfo {author} {\bibfnamefont {A.}~\bibnamefont {Ninarello}}, \ and\
  \bibinfo {author} {\bibfnamefont {M.}~\bibnamefont {Ozawa}},\ }\href@noop {}
  {\bibfield  {journal} {\bibinfo  {journal} {Physical Review Letters}\
  }\textbf {\bibinfo {volume} {116}},\ \bibinfo {pages} {238002} (\bibinfo
  {year} {2016})}\BibitemShut {NoStop}%
\bibitem [{\citenamefont {Katira}\ \emph {et~al.}(2018)\citenamefont {Katira},
  \citenamefont {Garrahan},\ and\ \citenamefont
  {Mandadapu}}]{katira2018solvation}%
  \BibitemOpen
  \bibfield  {author} {\bibinfo {author} {\bibfnamefont {S.}~\bibnamefont
  {Katira}}, \bibinfo {author} {\bibfnamefont {J.~P.}\ \bibnamefont
  {Garrahan}}, \ and\ \bibinfo {author} {\bibfnamefont {K.~K.}\ \bibnamefont
  {Mandadapu}},\ }\href@noop {} {\bibfield  {journal} {\bibinfo  {journal}
  {Physical Review Letters}\ }\textbf {\bibinfo {volume} {120}},\ \bibinfo
  {pages} {260602} (\bibinfo {year} {2018})}\BibitemShut {NoStop}%
\end{thebibliography}
\end{document}